\documentclass[trackchanges]{aastex701}
\usepackage{multirow}
\usepackage{xcolor}
\usepackage[normalem]{ulem}

\begin{document}

\title{Development of Electroformed X-ray Optics Bridging Synchrotron Technology and Space Astronomy}

\author[orcid=0009-0004-5957-3505]{Ryuto Fujii} 
\affiliation{Graduate School of Science, Tokai National Higher Education and Research System, Nagoya University, Furo-cho, Chikusa-ku, Nagoya, Aichi 464-8602, Japan}
\email{fujii_r@u.phys.nagoya-u.ac.jp}  

\author[orcid=0009-0005-2657-2554]{Koki Sakuta} 
\affiliation{Graduate School of Science, Tokai National Higher Education and Research System, Nagoya University, Furo-cho, Chikusa-ku, Nagoya, Aichi 464-8602, Japan}
\email{sakuta_k@u.phys.nagoya-u.ac.jp}  

\author[orcid=0009-0005-3295-7215]{Kazuki Ampuku} 
\affiliation{Graduate School of Science, Tokai National Higher Education and Research System, Nagoya University, Furo-cho, Chikusa-ku, Nagoya, Aichi 464-8602, Japan}
\email{ampuku_k@u.phys.nagoya-u.ac.jp}  

\author[orcid=0009-0005-2776-7775]{Yusuke Yoshida} 
\affiliation{Graduate School of Science, Tokai National Higher Education and Research System, Nagoya University, Furo-cho, Chikusa-ku, Nagoya, Aichi 464-8602, Japan}
\email{yoshida_y@u.phys.nagoya-u.ac.jp}  

\author[orcid=0009-0007-1256-4611]{Makoto Yoshihara} 
\affiliation{Graduate School of Science, Tokai National Higher Education and Research System, Nagoya University, Furo-cho, Chikusa-ku, Nagoya, Aichi 464-8602, Japan}
\email{yoshihara_m@u.phys.nagoya-u.ac.jp}

\author[]{Ayumu Takigawa} 
\affiliation{Graduate School of Science, Tokai National Higher Education and Research System, Nagoya University, Furo-cho, Chikusa-ku, Nagoya, Aichi 464-8602, Japan}
\email{a.takigawa3030@gmail.com}

\author{Keitoku Yoshihira} 
\affiliation{Graduate School of Science, Tokai National Higher Education and Research System, Nagoya University, Furo-cho, Chikusa-ku, Nagoya, Aichi 464-8602, Japan}
\email{yoshihira_k@u.phys.nagoya-u.ac.jp}  

\author{Tetsuo Kano}
\affiliation{Graduate School of Science, Tokai National Higher Education and Research System, Nagoya University, Furo-cho, Chikusa-ku, Nagoya, Aichi 464-8602, Japan}
\email{kano.tetsuo.s6@f.mail.nagoya-u.ac.jp}  

\author{Naoki Ishida} 
\affiliation{Graduate School of Science, Tokai National Higher Education and Research System, Nagoya University, Furo-cho, Chikusa-ku, Nagoya, Aichi 464-8602, Japan}
\email{ishida.naoki.i2@f.mail.nagoya-u.ac.jp}  

\author[orcid=0000-0002-6330-3944]{Noriyuki Narukage} 
\affiliation{National Astronomical Observatory Japan, 2-21-1 Osawa, Mitaka, Tokyo 181-8588, Japan}
\email{noriyuki.narukage@nao.ac.jp}  

\author[orcid=0000-0000-0000-0000]{Keisuke Tamura} 
\affiliation{NASA/GSFC, 8800 Greenbelt Rd, Greenbelt, MD 20771, US}
\affiliation{Center for Space Sciences and Technology, University of Maryland, Baltimore County (UMBC), Baltimore, MD, 21250 USA}
\email{ktamura1@umbc.edu}  

\author[orcid=0000-0000-0000-0000]{Kikuko Miyata} 
\affiliation{Meijo University, 1-501 Shiogamaguchi, Tempaku-ku, Nagoya, Aichi 468-8502, Japan}
\email{kmiyata@meijo-u.ac.jp}  

\author[orcid=0000-0003-4305-807X]{Gota Yamaguchi} 
\affiliation{RIKEN/ SPring-8, 1-1-1, Kouto, Sayo-cho, Sayo-gun, Hyogo 679-5198, Japan}
\email{g-yamaguchi@spring8.or.jp}  

\author[orcid=0000-0002-6841-0116]{Hidekazu Takano} 
\affiliation{RIKEN/ SPring-8, 1-1-1, Kouto, Sayo-cho, Sayo-gun, Hyogo 679-5198, Japan}
\email{h.takano@spring8.or.jp}  

\author[orcid=0000-0001-5182-063X]{Yoshiki Kohmura} 
\affiliation{RIKEN/ SPring-8, 1-1-1, Kouto, Sayo-cho, Sayo-gun, Hyogo 679-5198, Japan}
\email{yoshiki.kohmura@riken.jp}  

\author{Shutaro Mohri} 
\affiliation{School of Engineering, The University of Tokyo, 7-3-1 Hongo, Bunkyo, Tokyo 113-8656, Japan}
\email{mohri-shutaro644@g.ecc.u-tokyo.ac.jp}  

\author[orcid=0000-0002-1254-4297]{Takehiro Kume} 
\affiliation{Natsume Optical Corporation, 3461 Kamichyaya, Kanae, Iida, Nagano 395-0808, Japan}
\email{takehiro.kume@natsume-optics.co.jp}  

\author[orcid=0000-0002-2409-3956]{Yusuke Matsuzawa} 
\affiliation{Natsume Optical Corporation, 3461 Kamichyaya, Kanae, Iida, Nagano 395-0808, Japan}
\email{yusuke.matsuzawa@natsume-optics.co.jp}  

\author[orcid=0009-0000-9348-9164]{Yoichi Imamura} 
\affiliation{Natsume Optical Corporation, 3461 Kamichyaya, Kanae, Iida, Nagano 395-0808, Japan}
\email{yoichi.imamura@natsume-optics.co.jp}  

\author{Takahiro Saito} 
\affiliation{Natsume Optical Corporation, 3461 Kamichyaya, Kanae, Iida, Nagano 395-0808, Japan}
\email{takahiro.saito@natsume-optics.co.jp}  

\author[orcid=0000-0003-2231-3502]{Kentaro Hiraguri} 
\affiliation{Natsume Optical Corporation, 3461 Kamichyaya, Kanae, Iida, Nagano 395-0808, Japan}
\email{kentaro.hiraguri@natsume-optics.co.jp}  

\author[orcid=0000-0002-4589-0460]{Hirokazu Hashizume} 
\affiliation{Natsume Optical Corporation, 3461 Kamichyaya, Kanae, Iida, Nagano 395-0808, Japan}
\email{hirokazu.hashizume@natsume-optics.co.jp}  

\author[orcid=0000-0003-4623-6951]{Hidekazu Mimura} 
\affiliation{RIKEN/ SPring-8, 1-1-1, Kouto, Sayo-cho, Sayo-gun, Hyogo 679-5198, Japan}
\affiliation{Research Center for Advanced Science and Technology, The University of Tokyo, 4-6-1 Komaba, Meguro-ku, Tokyo 153-8904, Japan}
\email{mimura@upm.rcast.u-tokyo.ac.jp}  

\author[orcid=0000-0002-9901-233X]{Ikuyuki Mitsuishi} 
\affiliation{Graduate School of Science, Tokai National Higher Education and Research System, Nagoya University, Furo-cho, Chikusa-ku, Nagoya, Aichi 464-8602, Japan}
\email[show]{mitsuisi@u.phys.nagoya-u.ac.jp}  



\begin{abstract}

We have developed X-ray telescope mirrors using our original electroforming replication technique, which was established through the fabrication of millimeter-aperture, ultra-short-focal-length nanofocusing mirrors used in synchrotron X-ray microscopy.
%
In this paper, we present for the first time the detailed results of X-ray illumination tests performed both for a 60-mm-diameter, full-circumference, double-reflection monolithic electroformed nickel mirror and for the Mirror Module Assembly (MMA) incorporating it. 
The experiments were performed at SPring-8 beamline BL29XUL, which has a 1-km beamline length. 
To simulate the parallel X-ray beam from celestial sources, we constructed and utilized a dedicated evaluation system, the High-Brilliance X-ray Kilometer-long Large-Area Expanded-beam Evaluation System (HBX-KLAEES). 
Thanks to the high photon flux and the virtually point-like source with a very small diverging angle provided by HBX-KLAEES, we were able to evaluate the imaging performance with high fidelity, resolving both the sharp core and the large-angle components of the Point Spread Function (PSF).
The results demonstrated an extremely sharp core with a Full Width at Half Maximum (FWHM) of 0.7 arcsec and a Half Power Diameter (HPD) of 14 arcsec even after integration into the MMA.
Furthermore, a positive correlation was found between the angular resolution and the axial direction figure error of both the primary and secondary sections, indicating that the axial figure error contributes to image degradation. 
Based on these achievements, the MMA incorporating this mirror was selected as one of the hard X-ray optics for the FOXSI-4 sounding rocket experiment, which performs high-resolution soft and hard X-ray imaging spectroscopy of solar flares, and was successfully launched.
These results pave the way for further improvements in angular resolution and for the development of high-resolution, ultra-short focal length X-ray optics applicable to small satellites including CubeSats.

\end{abstract}


\section{Introduction} \label{sec:Intro}
Since the first discovery of an extrasolar X-ray source, space-based X-ray observations targeting high-energy phenomena associated with celestial objects--such as black holes, stars, supernova remnants, galaxies, galaxy clusters, and large-scale structure of the Universe--have revealed the dynamic nature of the universe.
From an instrumentation perspective, the Einstein Observatory achieved the first imaging observations with focusing X-ray optics, representing one of the key milestones in the history of X-ray astronomy\citep[e.g.,][]{Giacconi1979apj}. 
Since then, X-ray optics have been mounted as primary instruments on many balloons, sounding rockets, and satellites, playing essential roles in achieving scientific goals and pioneering new frontiers in high-energy astrophysics\citep[e.g.,][]{Aschenbach1985rpp,Christensen2022nature,Wilkes2022nature}. 

Among the key performance metrics of space-based X-ray optics, angular resolution and effective area are of primary importance. 
High angular resolution allows for detailed imaging of extended celestial sources, enabling the study of the spatial distributions of plasma properties such as density and temperature. 
For compact sources, it reduces contamination from nearby emissions in crowded fields --including star clusters and the Galactic center-- enabling higher-quality imaging and spectroscopy. 
Improved angular resolution also enhances the signal-to-noise ratio by suppressing detector noise and diffuse X-ray background, thus contributing to higher detection sensitivity.
The angular resolution of X-ray optics is usually characterized by the Full Width at Half Maximum (FWHM) of the Point Spread Function (PSF) and the Half Power Diameter (HPD) in X-ray astronomy.
FWHM reflects the sharpness of the PSF core and indicates the ability to separate closely spaced point sources. In contrast, HPD represents the diameter of the circle enclosing 50\% of the total flux and accounts for components scattered to relatively large angles, which are relevant for observations of extended sources. 

A wide range of fabrication techniques have been explored in the development of X-ray optics -- for example, polishing of glass; replication using nickel, nickel--cobalt, aluminum, or CFRP; and multilayer coatings, including depth-graded multilayers (so-called supermirrors) based on atomic layer deposition (ALD) or sputtering technologies\citep[e.g.,][]{Weisskopf2012spie,Aschenbach2002spie,Ramsey2022jatis,Serlemitsos2007pasj,Awaki2022spie,Chan2013spie,Awaki2014apopt}. 
In addition, alternative optical designs, such as micro-pore optics and interferometric approaches, have also been developed\citep[e.g.,][]{Mitsuishi2010apopt,Numazawa2024spie,Collon2015spie,Zhang2019jatis,Li2024optl,Kitamoto2014spie,Asakura2024jatis}. 
Particularly for large X-ray optics aboard large astronomical satellites (e.g., Chandra, XMM-Newton, and XRISM), which typically require focal lengths of several meters or more and diameters of tens of centimeters or larger, significant progress has been made in specialized fabrication technologies dedicated to these unique design requirements. 
Meanwhile, ground-based technologies have often served as key drivers for advancing space-based X-ray optics.

Building on these developments, we focused on electroforming technology to realize high--angular-resolution space X-ray optics because it enables the fabrication of double-reflection, full-circumference monolithic mirrors, which allows for precise alignment compared with segmented mirrors, both between the primary and secondary stages and among mirrors in the circumferential direction.
In addition, once a high-quality mandrel is prepared, the same mandrel can be reused to produce multiple mirrors cost-effectively. 
For these reasons, we adopted our original electroforming technology, which is based on techniques developed and continually refined through the fabrication of millimeter-aperture, ultra-short-focal-length nanofocusing mirrors used in synchrotron X-ray microscopy\citep[e.g.,][]{Mimura2018rsi,Kume2019rsi,Takeo2020apphl}.
%
%
Our electroforming process is characterized by non-agitated, room-temperature deposition, which achieves nanometer-level surface transfer accuracy for small mirrors with $\sim$10 mm aperture and focal lengths below 100 mm. 
The fabricated optics have reached nearly diffraction-limited performance, with spot sizes approaching the theoretical limit in terms of FWHM, mainly due to the small circumferential figure errors. 

We thus extended this electroforming process to larger-diameter telescope mirrors aimed at space-based X-ray optics. 
As a first step, we increased the diameter from 10 mm to 60 mm and optimized both the machining and polishing of the mandrel and the design and fabrication of the electroforming setup. 
In our previous study, we successfully fabricated a Wolter-I--type mirror shell using a quartz-glass mandrel\citep{Yamaguchi2020spie,Ito2021spieproc,Yamaguchi2023rsi}. 
Although the formation of pits due to trapped gas bubbles had been a major obstacle in producing large-diameter electroformed nickel mirrors, we implemented a vacuum degassing process to suppress pit formation effectively, achieving pit-free and highly precise replication. 
The resulting mirror showed figure errors below 1 $\mu$m over most of both the axial and circumferential directions. 
A ray-tracing simulation based on these measured figure errors predicted an HPD of approximately 12 arcsec\citep{Yamaguchi2023rsi}.

For the first application of this technology, we selected the Sun as the observational target, where a small effective area is acceptable but high imaging capability is essential. 
The Focusing Optics X-ray Solar Imager (FOXSI) sounding rocket experiment is a Japan--US collaborative mission that aims to investigate both thermal and nonthermal processes in the solar corona and flares through the world’s first soft and hard X-ray imaging spectroscopy\citep[e.g.,][]{Krucker2013spie,Christe2016jai,Buitrago-Casas2022AandA,Glesener2023bulletin}. 
In particular, FOXSI-4, the fourth flight of the series, was designed to achieve the highest angular resolution among the FOXSI series and to perform the first high-resolution soft and hard X-ray imaging spectroscopy of solar flares. 
Once the fabrication framework for this optics is established, it will pave the way toward the realization of ultra-short focal length ($\sim$100 mm) and high-angular-resolution X-ray optics -- an essential technology for small satellite missions including CubeSats that are now regarded as potential game changers in space science.

In this paper, we present the first detailed X-ray performance evaluation of a 60-mm-diameter, full-circumference, double-reflection monolithic nickel mirror fabricated by our electroforming process, as well as of the Mirror Module Assembly (MMA) incorporating it, with a focus on the achieved angular resolution. 
We also describe the dedicated evaluation system constructed at SPring-8 beamline BL29XUL for detailed performance testing.
The structure of this paper is as follows: Section~2 introduces the fabricated samples; Section~3 presents the experimental setup, procedures, and results of the X-ray illumination tests; Section~4 discusses the origin of angular resolution degradation; and Section~5 provides a summary and conclusions.

Throughout this paper, all data analyses were performed using Python 3.9.22 and standard scientific software packages, including Astropy 6.0.1 \citep{astropy:2013, astropy:2018, astropy:2022}, NumPy 1.26.4 \citep{harris2020array}, SciPy 1.13.1 \citep{2020SciPy-NMeth},
Matplotlib 3.9.4 \citep{Hunter:2007}, and pandas 2.2.3 \citep{reback2020pandas}.

\section{Sample} \label{chap_sample}
This section describes the X-ray mirror fabricated using our original electroforming technology, which was established through the fabrication of millimeter-aperture, ultra-short-focal-length nanofocusing mirrors used in synchrotron X-ray microscopy, as well as the structure of the MMA into which the mirror was integrated.

\paragraph{Mirror}

The Wolter-I type mirror designed and fabricated in this study has the following parameters: a focal length of 2 m, a diameter of 60 mm, a total effective length of 200 mm (primary: 102.5 mm; secondary: 97.5 mm), an incidence angle of 0.21${}^\circ$, a thickness of 2 mm, and a mass of 760 g.
In this grazing-incidence optical configuration, basically, X-rays are reflected twice--first by the primary (paraboloid) and then by the secondary (hyperboloid)--to form an image. 
The focal length is defined as the distance from the intersection of the two sections to the focal plane. 
The greatest technical challenge in X-ray mirror fabrication lies in achieving larger diameters. 
As a first step, we adopted a 60 mm-diameter mirror, approximately six times larger than our previous prototypes. 
The critical energy for this geometry is 16 keV. 
The heights of the primary and secondary sections were determined so that photons incident along the optical axis that are reflected by the primary surface are subsequently reflected by the secondary surface, thereby maximizing the effective area. 
No coating was applied to the mirror surface. 
These design parameters were determined by considering both the scientific requirements and technical constraints for solar-flare observations, for which high angular resolution was prioritized over effective area.

\paragraph{Mirror Module Assembly (MMA)}

\begin{figure*}[h!]
  \centering
  \includegraphics[width=12cm]{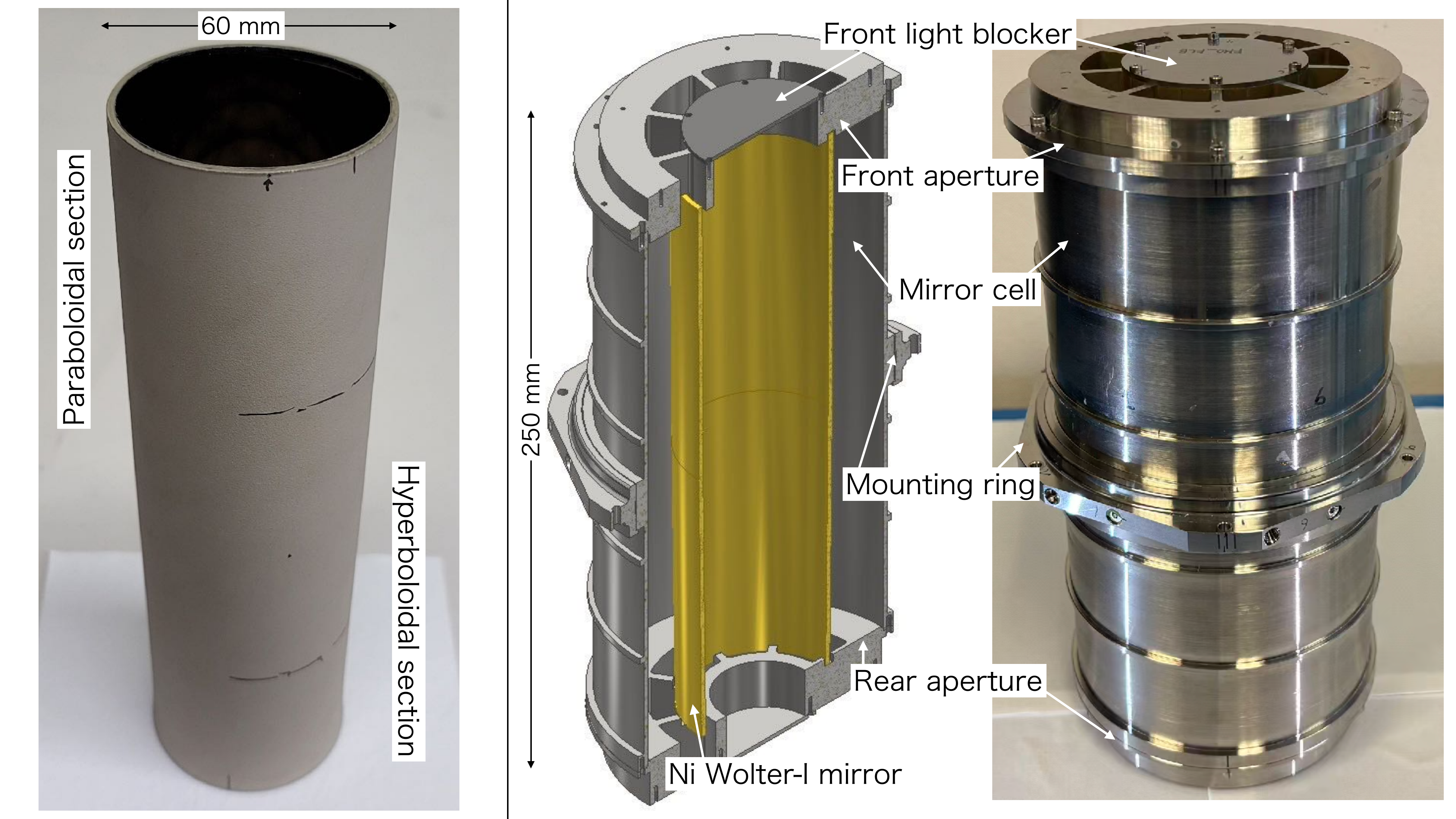}
  \caption{Electroformed X-ray mirror (left), schematic of the MMA design (center), and photograph of the MMA (right).}
  \label{fig:mirror_mma}
\end{figure*}

The MMA consists of front and rear apertures, the mirror itself, a mirror support structure, a sounding rocket interface, and front and rear light blockers. 
Its basic design follows the structure used in previous FOXSI missions\citep{Christe2016jai}, with all components, except for the mirror, made of stainless steel. 
The MMA is designed to withstand the mechanical loads experienced during sounding rocket launch. 
In this design, the mirror is held only by the rear aperture to preserve its intrinsic shape. 
The aperture includes nine radially arranged spokes, each containing a groove for positioning and bonding the mirror with adhesive. 
The focal length reference is aligned with the rocket mounting interface. 
To prevent stray light from directly reaching the detector without reflection, front and rear light blockers were installed; their thicknesses were chosen to ensure sufficient opacity at the critical energy. 
After successful X-ray performance evaluation (described in the following section), this MMA was selected as one of the hard X-ray optics for the FOXSI-4 sounding rocket mission\citep{Ampuku2024spie}, in conjunction with a hard X-ray focal plane detector covering the energy range from 4 to 20 keV\citep{Nagasawa2025nima}. 
Figure \ref{fig:mirror_mma} shows photographs of the fabricated electroformed mirror and the MMA, together with a schematic design of the MMA.

\section{X-ray Illumination Tests} \label{sec:xraytests}
\subsection{Experimental Setup}
We conducted X-ray illumination tests at the synchrotron radiation facility SPring-8. At SPring-8, electrons are accelerated to high energies and deflected by magnetic fields, generating synchrotron radiation, i.e., highly collimated and intense X-rays. 
SPring-8 can accelerate electrons up to 8 GeV, placing it among the world’s few third-generation light sources capable of providing extremely high brilliance and monochromaticity. 
Compared with conventional fluorescent and bremsstrahlung X-rays produced by irradiating metal targets with thermal electrons, synchrotron radiation enables far superior spectral purity and flexible energy selection. 
Moreover, the extremely high brilliance and stability of the beam enable precise characterization of both the core and large-angle scattering components within a short exposure time, even for optics with small effective areas. 
SPring-8 is equipped with 57 beamlines covering a broad spectral range from soft X-rays at 170 eV to hard X-rays up to 300 keV, enabling selection of the most appropriate beamline for each experimental purpose.

\begin{figure*}[h!]
  \centering
  \includegraphics[width=15cm]{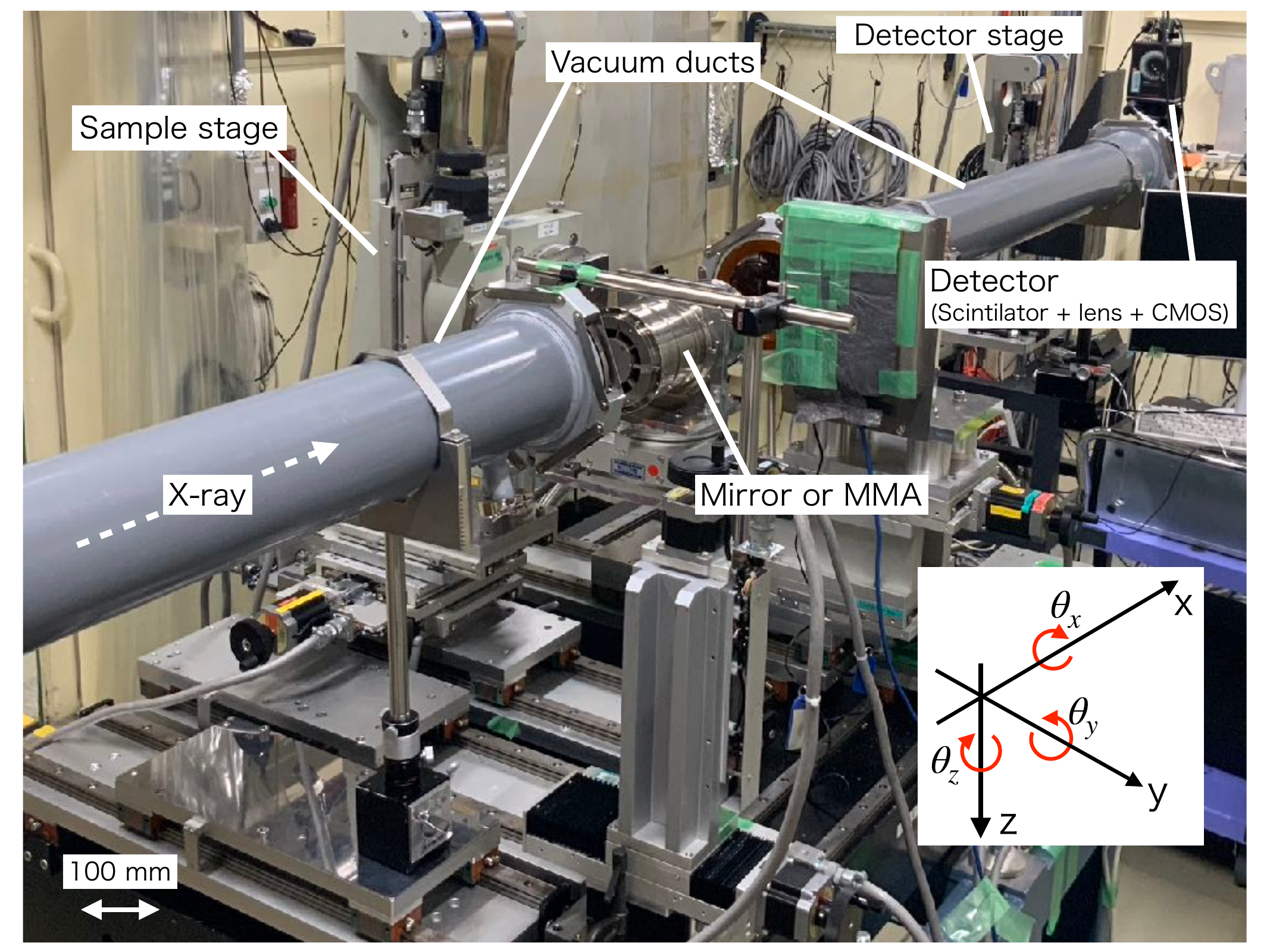}
  \caption{Experimental setup inside the fourth experimental hutch of SPring-8 beamline BL29XUL.
The arrow indicates the direction of incident X-ray photons from the upstream side.
The setup consists of a vacuum duct, sample, sample stage, detector, and detector stage.
The coordinate axes are also defined in the figure.}
  \label{fig:setup_sp8BL29XUL}
\end{figure*}

For this work, we employed BL29XUL\citep{Ishikawa2001spie,Tamasaku2001nima}, the beamline with the longest optical path at SPring-8--approximately 1 km in length. 
Such an extended beamline is ideal for simulating parallel X-rays arriving from astronomical sources at infinity, which is essential for high-precision characterization of space X-ray optics. 
The peak energy of the beamline was approximately 10 keV, with a tunable range between 5--38 keV using standard in-vacuum undulator and double crystal monochromator (DCM). 
A schematic of the experimental setup inside the fourth experimental hutch is shown in Figure \ref{fig:setup_sp8BL29XUL}. X-rays entered from the left (upstream side), passing sequentially through the vacuum duct -- mirror sample -- vacuum duct -- detector. 
To minimize scattering by air, the ducts were evacuated to a pressure of slightly lower than 10 Pa. 
Measurements were performed at a representative energy of 12 keV, which lies near the middle of the observational energy band when combined with the focal-plane detector.
At this energy, the influence of air scattering on large-angle components was confirmed to be negligible.
The coordinate system for the sample and detector stages is defined as follows: the x-axis is along the beam direction (corresponding to the focal length direction), the y-axis is transverse to the beam in the horizontal plane, and the z-axis is vertical.
The rotation angles $\theta_{\rm{x}}$, $\theta_{\rm{y}}$, and $\theta_{\rm{z}}$ represent rotations about each corresponding axis. 
The detector was mounted on an automated translation stage movable in the y--z plane, while the mirror stage provided five degrees of freedom: x, y, z, $\theta_{\rm{y}}$, and $\theta_{\rm{z}}$.
Two types of measurements were performed:
\begin{enumerate}
  \item Spot-scan test -- evaluation of the local imaging performance of the mirror alone by scanning a small-area beam along the mirror surface. 
  \item Full-illumination test -- evaluation of the overall imaging performance of the MMA using a large, expanded X-ray beam that illuminated the entire aperture.
\end{enumerate}
The following subsections describe each method in detail.

\subsection{Spot-scan Method}
To evaluate the spatial dependence of the mirror’s local angular resolution, we used an X-ray beam with a size of 5 × 5 mm$^{2}$ and irradiated 40 positions covering the entire mirror. 
During each measurement, the beam position was fixed while the mirror and detector stages were synchronously translated to maintain their relative alignment. 
At each position, an image was acquired at the best focus. The beam uniformity in the illuminated area was measured to be 10 \% (RMS).



\subsection{Full-illumination Method (HBX-KLAEES)}
\begin{figure*}[h!]
  \centering
  \includegraphics[width=15cm]{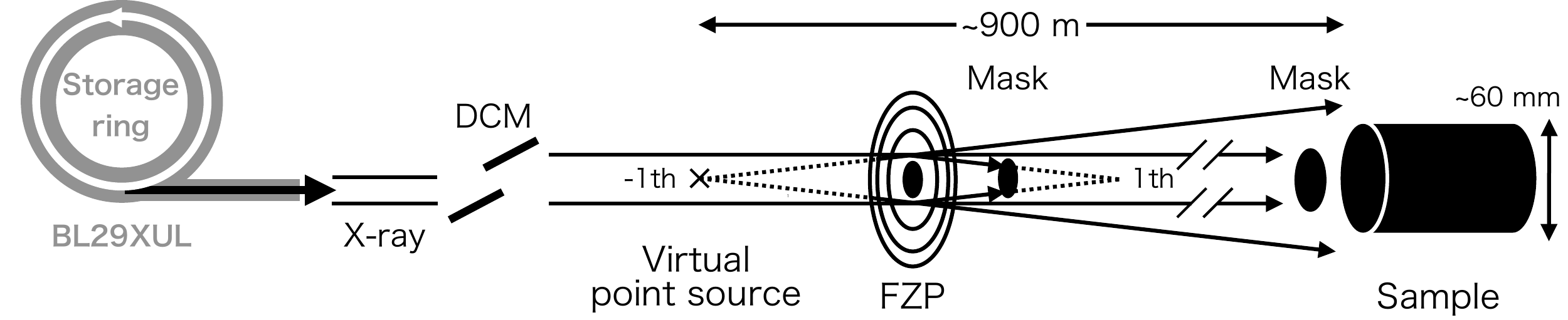}
  \caption{Conceptual diagram of the High-Brilliance X-ray Kilometer-long Large-Area Expanded-beam Evaluation System (HBX-KLAEES) constructed at SPring-8 beamline BL29XUL.
  %
  The distance from the virtual point source to the sample is approximately 1 km, enabling highly collimated beam conditions.}
  \label{fig:hbx-klaees}
\end{figure*}

In previous ground-based evaluations for Japanese X-ray astronomy missions, the overall performance of the optics was sometimes assessed using a raster-scan method\citep[e.g.,][]{Shibata2001ao,Itoh2004spieproc,Hayashi2015jatis,Iizuka2018jatis}, in which a small beam was scanned across the aperture and the images were combined to reconstruct the full image. 
However, for high-angular-resolution optics, the achievable accuracy is limited by the synchronization precision between stages, making the method unsuitable for arcsecond measurements. 
To overcome this limitation, we developed and utilized a dedicated system--the High-Brilliance X-ray Kilometer-long Large-Area Expanded-beam Evaluation System (HBX-KLAEES)--at SPring-8 beamline BL29XUL. 
The system was designed to simulate parallel X-rays from astronomical sources by producing a wide and nearly collimated beam over the full mirror aperture. 
Since standard SPring-8 setups are optimized for millimeter-scale samples, a specialized optical arrangement was required. 
We introduced a Fresnel Zone Plate (FZP) upstream of the beamline to generate a virtual point source. 
With the 1 km propagation distance of BL29XUL, the diverging beam from the FZP becomes effectively parallel at the mirror position, achieving both high collimation and sufficient beam coverage for full-aperture illumination. 
The FZP (zone material: Au, zone thickness: 1500 nm, diameter: 1 mm, outermost zone width: 1000 nm; Applied Nanotools Inc., Canada) was located 95 m from the source, with multiple diffraction orders; the --1st order (divergent component with the highest intensity) was used. 
At 12 keV, the focal length of the FZP was 9.67 m. The mirror sample was positioned 988 m from the source, corresponding to a virtual source-to-sample distance of 904 m. 
The focusing size of the +1st-order beam was experimentally measured to be 23 $\mu$m (vertical) and 44 $\mu$m (horizontal) in FWHM at 6 keV with a focal length of 4.8 m. 
The -1st-order beam is expected to have a similar focus size, and at our measurement energy of 12 keV, these values are expected to increase by approximately a factor of two due to the energy dependence of diffraction. 
Consequently, when viewed from the sample, the apparent source size remains sufficiently small ($\sim$0.02 arcsec), and the diverging angle is also small ($\sim$7 arcsec). 
These conditions enable precise measurements of both on- and off-axis responses in terms of high-angular-resolution performance and fundamental characteristics such as effective area, field of view, and depth of focus.
%
The FZP aperture was 1 mm, producing a theoretical beam diameter of 93 mm, sufficient to cover the mirror’s 60 mm aperture.
We confirmed that the actual beam profile--slightly distorted by pipe curvature and transport effects--still provided full coverage ($\sim$ 66 mm diameter). 
To ensure that only the divergent -1st order beam illuminated the sample, the convergent +1st order component was blocked by a mask placed at its focal position, and the direct beam was also blocked by another mask positioned just in front of the sample.
%
The beam uniformity at the mirror aperture was approximately 20 \% (RMS). A conceptual diagram of HBX-KLAEES is shown in Figure \ref{fig:hbx-klaees}.

\subsection{Detector System}
The imaging detector consisted of a scintillator-coupled CMOS camera (indirect conversion system) by Hamamatsu Photonics, Japan. 
A relay lens between the scintillator (LuAG: Ce, 5-$\mu$m-thick) and CMOS sensor allowed adjustment of the effective pixel scale. 
For the spot-scan test, the pixel size was 6.5 $\mu$m with a resolution of 2048 × 2048 pixels using ORCA-FLASH corresponding to a field of view (FOV) of 23 × 23 arcmin$^2$. 
For the full-illumination test, we used a higher-resolution configuration with 4.5 $\mu$m pixels and 2304 × 4096 pixels, using ORCA-Quest, yielding a FOV of 18 × 32 arcmin$^2$. 
The scintillator had a diameter of 10 mm, defining the effective FOV; this corresponded to a full field of view of approximately 17 arcmin.

\section{Imaging Performance} \label{sec:results}
\subsection{Alignment of Optical Axis and Focal Adjustment}

Before evaluating the angular resolution, the optical axis and focal position of the sample were carefully aligned. 
For optical-axis alignment, the incidence angle to the sample was varied to locate the position of maximum effective area. 
%
The alignment was performed sequentially around the two rotation axes, $\theta_{\rm{y}}$ and $\theta_{\rm{z}}$.
The off-axis dependence of the effective area was measured and fitted to determine the peak position. 
%
%
%
The focus position was then adjusted. 
%
%
The position yielding the best imaging performance was defined as the best focus. 
For the MMA, however, a finer focus scan was performed in consideration of the mounting accuracy during sounding rocket integration, and the depth of focus--where the apparent imaging performance remained nearly unchanged--was also evaluated. 
The measured depth of focus was approximately 3 mm, and its midpoint was defined as the best-focus position.
%
%
The measured best-focus distances in both tests were consistent with the design value within the measurement uncertainty.

After the optical axis and focal position were aligned as described above, image sequences were acquired under the following conditions.
During the spot-scan test, 200 images were obtained with an exposure time of 5 ms per frame, while during the full-illumination test, 1000 images were taken with an exposure of 30 ms each.
All images were summed after background subtraction to produce a single composite image for analysis.
The subsequent analysis procedures are described in Section \ref{subsec:analysis-results}.

\subsection{Analysis and Results}\label{subsec:analysis-results}
After the alignment and focusing procedures described above, the acquired images were analyzed to evaluate the imaging performance of the mirror and the MMA using the spot-scan and full-illumination methods, respectively.
For background correction, a dark frame was acquired by closing the upstream beam shutter under the same conditions, and this was subtracted from the illuminated images. 
In addition, due to the relatively low photon statistics in the spot-scan test, short-term variations of the background were also evaluated using source-free regions within the field of view and incorporated into the background estimation. 
Because a larger image area was analyzed in the full-illumination test, hot-pixel correction was applied only in this case, by replacing each hot pixel with the mean of its surrounding pixels.
The angular resolution was evaluated using the FWHM and HPD metrics. 
The PSF was obtained by extracting the radial intensity profile centered at the maximum pixel value, and the Encircled Energy Function (EEF) was calculated by integrating and normalizing this profile. 
The HPD was defined as twice the radius corresponding to 0.5 in EEF. During analysis, the extraction radius was carefully set to exclude direct X-rays that bypassed the mirror as well as stray light produced by single reflections from either the primary or secondary sections.
The results of these analyses are presented below for both the spot-scan and full-illumination tests.

\begin{figure*}[h!]
  \centering
  \includegraphics[width=18cm]{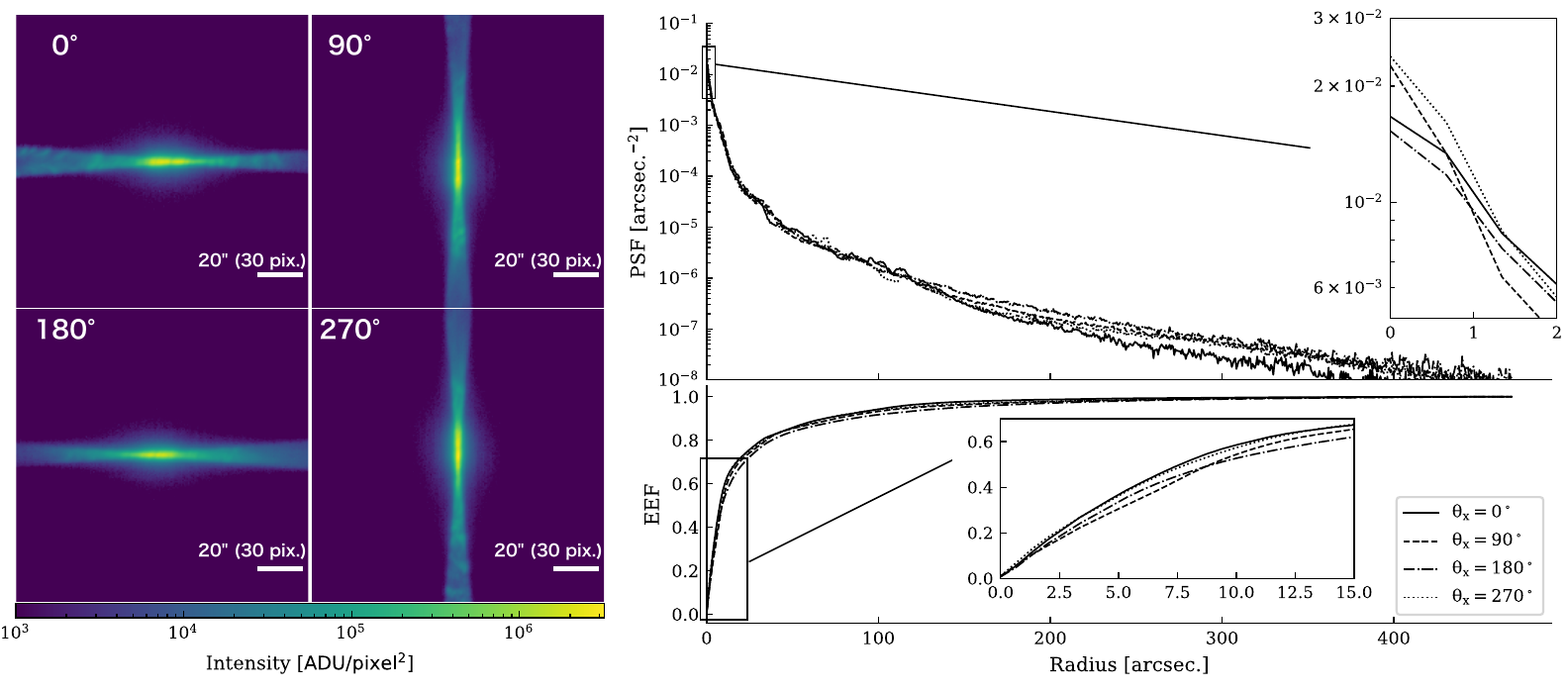}
 \caption{Left : example of X-ray images acquired from local illumination regions of the mirror using the spot-scan method at 12~keV. Images at phase angles of 0${}^\circ$ (upper left), 90${}^\circ$ (upper right), 180${}^\circ$ (lower left), and 270${}^\circ$ (lower right) are shown.
 Right : same as Figure left panels, but showing the PSF and EEF profiles.}
  \label{fig:spotscan_images-psf-eef}
\end{figure*}

\begin{figure*}[h!]
  \centering
  \includegraphics[width=9cm]{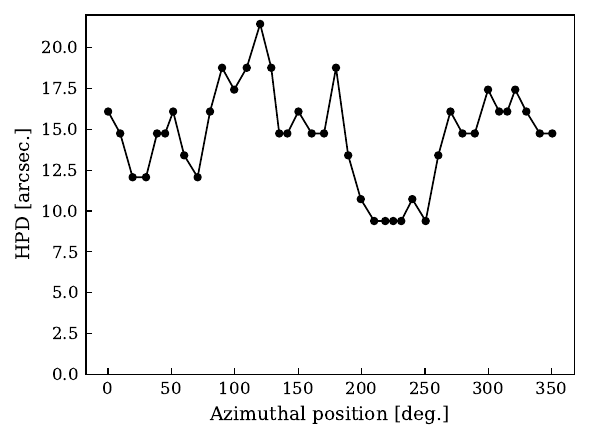}
  \caption{Azimuthal dependence of the angular resolution (HPD) of the mirror obtained using the spot-scan method at 12~keV.}
  \label{fig:spotscan_hpd}
\end{figure*}

\paragraph{Spot-scan Test of the Single Mirror}
After performing the alignment and focusing procedures described above, the mirror was first evaluated using the spot-scan method, placed on a V-block support in a single-mirror configuration before being integrated into the MMA.
In this test, the mirror was evaluated in 40 azimuthal regions around its circumference. Representative images, PSFs, and EEFs obtained from these measurements are shown in Figure \ref{fig:spotscan_images-psf-eef}. 
The extremely high photon statistics achieved in the synchrotron measurements enabled high-quality data that included scattering components more than five orders of magnitude below the PSF core, even for local illumination regions.
%
The reproducibility of the PSF and EEF profiles between two independent measurements was confirmed to be better than 10 and 1 \% (RMS), respectively. 
The measured FWHM values were approximately 3 arcsec (4 pixels) at all positions, showing an extremely sharp core. 
However, the measured FWHM was limited by the detector pixel size; thus, further FWHM analysis was conducted under higher magnification in the full-illumination test, where the optical axis was precisely aligned.
Here, we focus on the HPD values obtained for each azimuthal position, summarized in Table \ref{table:spotscan_hpd} and Figure \ref{fig:spotscan_hpd}. 
The HPD varied significantly with azimuth, ranging roughly from 10 to 20 arcsec, with a mean around 15 arcsec. 
For example, regions between 200${}^\circ$ and 250${}^\circ$ achieved HPDs of about 10 arcsec, while regions near 100${}^\circ$--150${}^\circ$ exceeded 20 arcsec.

%
%
%




\begin{table}[htbp]
\centering
\hfill
\begin{minipage}{0.9\textwidth}
\centering
\begin{tabular}{lcc|lcc|lcc} \hline
Position & HPD & slope error\footnote{Primary/Secondary sections\label{fot:a}} & Position & HPD & slope error\footref{fot:a} & Position & HPD & slope error\footref{fot:a} \\
{[deg.]} & {[arcsec.]} & {[arcsec.]} & {[deg.]} & {[arcsec.]} & {[arcsec.]} & {[deg.]} & {[arcsec.]} & {[arcsec.]} \\ \hline
  -5 - 4 & 16 & 2.2/2.9 & 125 - 132 & 18 & 1.5/2.2 & 245 - 255 & 9 & 1.0/1.1 \\
  4 - 14 & 14 & 1.6/2.4 & 132 - 138 & 14 & 1.7/2.2 & 255 - 265 & 13 & 1.2/1.3 \\
  14 - 24 & 12 & 1.7/1.7 & 138 - 144 & 14 & 1.8/2.1 & 265 - 274 & 16 & 1.4/1.6 \\
  24 - 35 & 12 & 1.7/1.2 & 144 - 155 & 16 & 1.7/1.9 & 274 - 284 & 14 & 1.5/1.7 \\
  35 - 41 & 14 & 1.7/1.9 & 155 - 165 & 14 & 1.4/1.7 & 284 - 294 & 14 & 1.4/1.7 \\
  41 - 48 & 14 & 1.7/2.1 & 165 - 175 & 14 & 1.1/1.4 & 294 - 305 & 17 & 1.3/2.1 \\
  48 - 54 & 16 & 1.6/1.9 & 175 - 184 & 18 & 0.9/1.2 & 305 - 311 & 16 & 1.1/2.3 \\
  54 - 65 & 13 & 1.4/1.9 & 184 - 194 & 13 & 0.8/1.3 & 311 - 318 & 16 & 1.2/2.4 \\
  65 - 75 & 12 & 1.1/1.8 & 194 - 204 & 10 & 0.9/1.3 & 318 - 324 & 17 & 1.2/2.7 \\
  75 - 85 & 16 & 0.8/1.7 & 204 - 215 & 9 & 0.9/1.1 & 324 - 335 & 16 & 1.1/2.9 \\
  85 - 94 & 18 & 1.2/1.9 & 215 - 222 & 9 & 0.8/1.1 & 335 - 345 & 14 & 1.1/3.0 \\
  94 - 104 & 17 & 1.4/2.0 & 222 - 228 & 9 & 0.8/1.2 & 345 - 355 & 14 & 1.4/3.0 \\
  104 - 114 & 18 & 1.5/2.3 & 228 - 234 & 9 & 0.8/1.2 & &&\\
  114 - 125 & 21 & 1.4/2.3 & 234 - 245 & 10 & 0.9/1.2 & &&\\
  \hline
\end{tabular}
\end{minipage}
\hfill
\caption{Illumination positions on the mirror during the spot-scan test, the angular resolution (HPD) measured at 12 keV, and the corresponding spread of the slope error distribution at each position.}
\label{table:spotscan_hpd}
\end{table}
\paragraph{Full-illumination Test for the MMA Using HBX-KLAEES}

\begin{figure*}[h!]
  \centering
  \includegraphics[width=18cm]{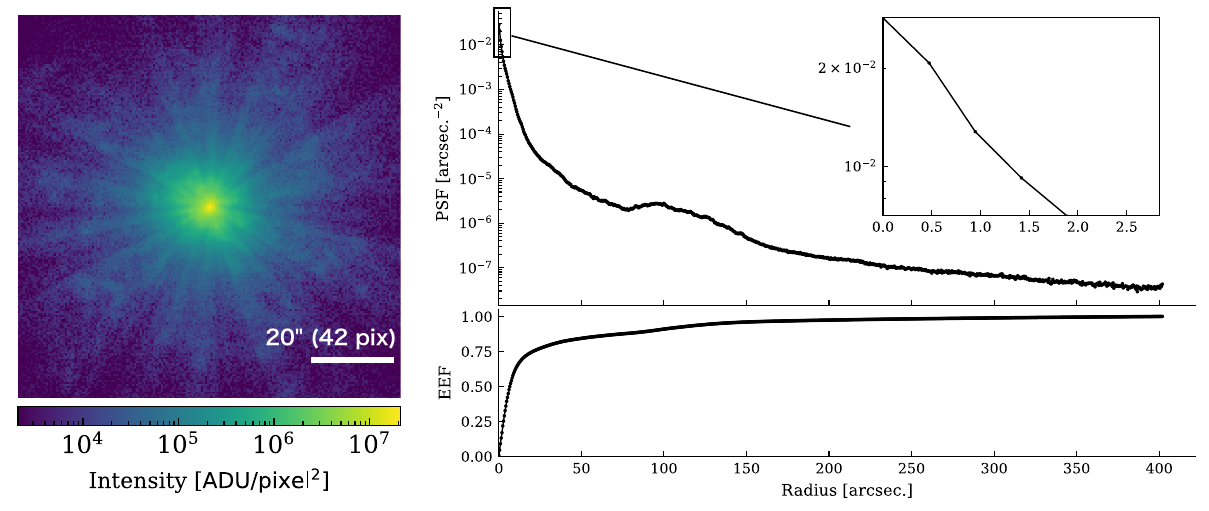}
  \caption{Full-aperture X-ray image of the MMA at 12 keV obtained using HBX-KLAEES.}
  \label{fig:full_image}
\end{figure*}

\begin{figure*}[h!]
  \centering
  \includegraphics[width=15cm]{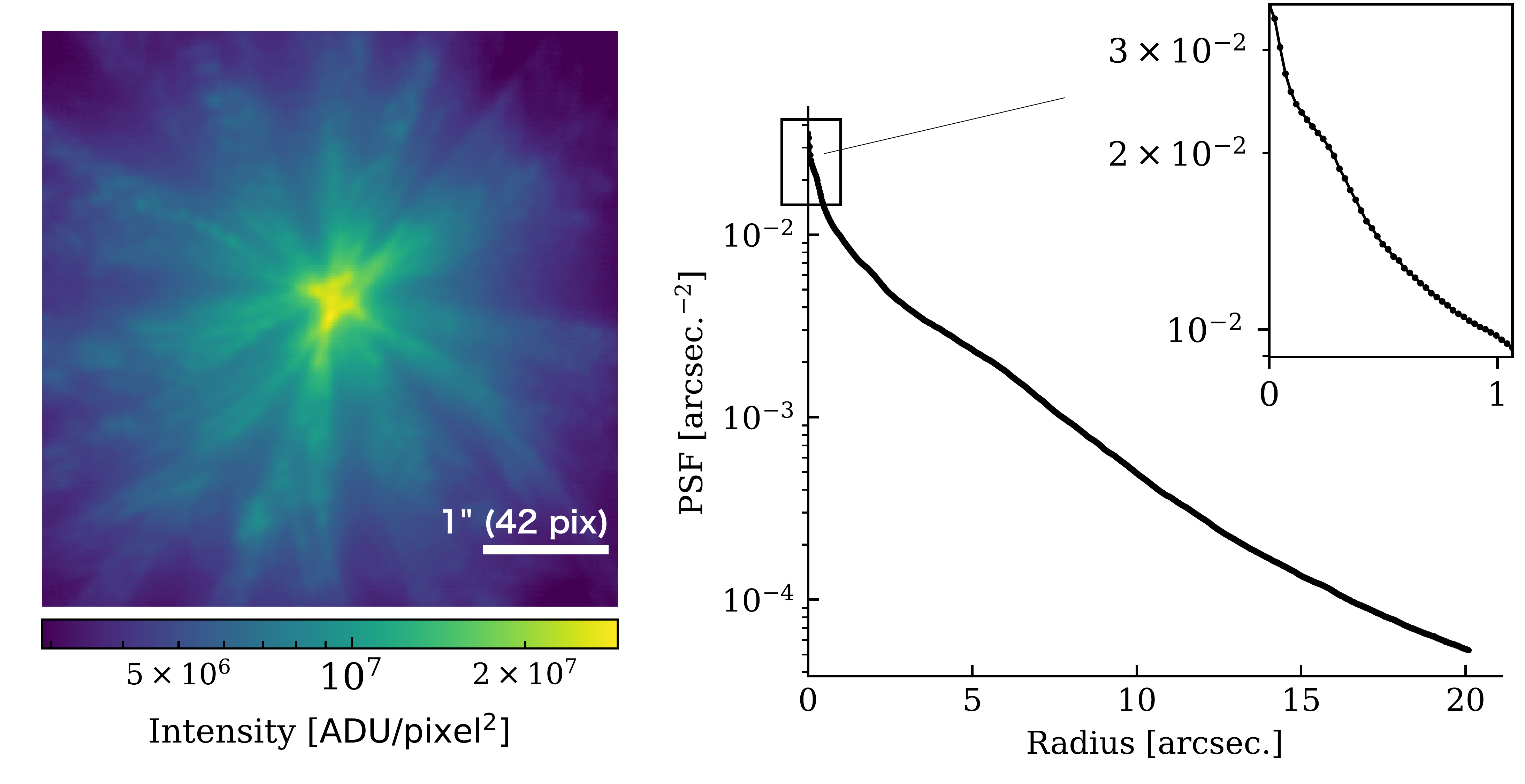}
  \caption{Same as Figure \ref{fig:full_image}, but acquired with a 20× lens magnification.}
  \label{fig:full_image-psf-eefx20}
\end{figure*}

Figure \ref{fig:full_image} shows the image, PSF, and EEF obtained in the full-illumination test using HBX-KLAEES. 
Similar to the spot-scan method, thanks to the extremely high photon flux of the SPring-8 beamline, we obtained high-quality data that included scattered components extending more than five orders of magnitude below the PSF peak intensity.
The reproducibility of the PSF and EEF profiles was confirmed within 20 and 1 \% (RMS), respectively. 
The measured FWHM and HPD were 3 arcsec (4 pixels) and 14 arcsec, respectively. 
The HPD value was almost consistent with the previous ray-tracing simulation that accounted for the measured figure errors\citep{Yamaguchi2023rsi}. 
The slight difference from the previous X-ray illumination results is attributed to the fact that, in the earlier study, only about half of the MMA was illuminated instead of the entire aperture, suggesting that the variation arises from position-dependent effects\citep{Fujii2024spie}.
Because the measured FWHM was determined from only a few detector pixels, it was limited by the camera resolution. 
To overcome this limitation, we increased the optical magnification from 1× to 20×, reducing the effective pixel size from 4.6 $\mu$m (0.47 arcsec) to 0.23 $\mu$m (0.023 arcsec). 
Figure \ref{fig:full_image-psf-eefx20} shows the resulting high-resolution image and PSF. 
The PSF was normalized to match the EEF obtained under the 1× configuration. 
Under this condition, the measured FWHM reached 0.7 arcsec (30 pixels), demonstrating an exceptionally sharp core. 
This result is attributed to the small azimuthal figure error and/or to the fact that the mirror is fabricated as a monolithic double-reflection mirror, which avoids misalignment between multiple mirrors or segments. 
Therefore, the MMA is expected to resolve very fine solar structures in bright X-ray emission regions. 
Moreover, HBX-KLAEES proved essential for accurate evaluation of such sub-arcsecond FWHM values. 

As the mirror had previously been confirmed to possess an extremely high degree of circularity in previous studies\citep{Yamaguchi2023rsi}, the PSFs obtained from the 40 spot-scan positions were co-added--assuming their peak pixel positions were aligned--and the resulting EEF gave an HPD of 15 arcsec, consistent within measurement uncertainty with the result from the full-illumination test.
%
This indicates that integration of the mirror into the MMA did not cause any measurable degradation of angular resolution. 
These results demonstrate that we have successfully established fundamental processes for mirror fabrication, evaluation, and mechanical integration into the MMA. 
Given the demonstrated performance and the process reproducibility, this technology represents a significant step toward realizing high-angular-resolution X-ray optics for both future astronomical missions and ultra-short-focal-length systems designed for small satellite applications including CubeSats.
%
%
Furthermore, our X-ray optics achieved higher angular resolution than those onboard the previous FOXSI series, marking an important milestone for the next generation of high-energy solar observations.

\section{Discussion} \label{sec:discussion}

The degradation of angular resolution in X-ray optics onboard previous missions has often been attributed, at least in part, to figure errors along the axial direction of the mirror surfaces. 
The extent of this degradation can be quantitatively evaluated by examining the width of the slope error histogram derived from the measured axial figure error profiles. 
In this section, we investigate the correlation between the X-ray performance acquired through the spot-scan measurements and the measured axial figure errors of the mirror surfaces, in order to clarify one of the possible causes of angular resolution degradation.

\begin{figure*}[h!]
  \centering
  \includegraphics[width=18cm]{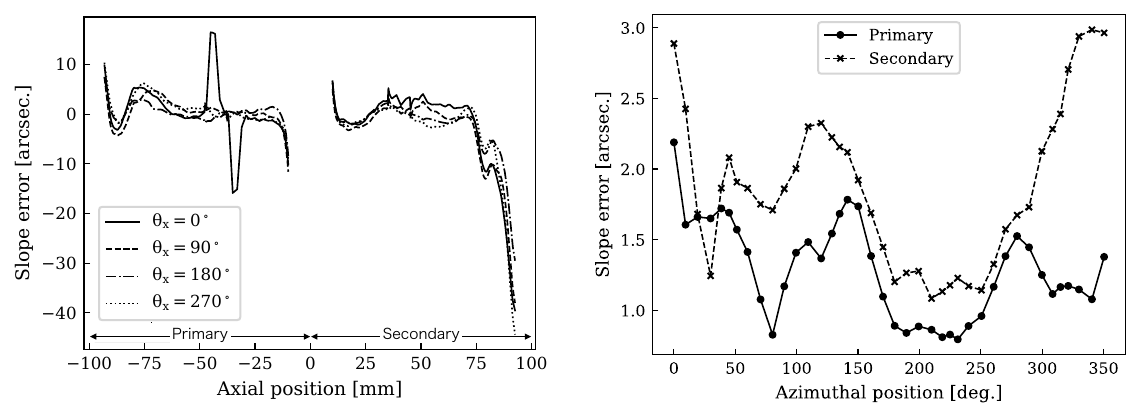}
  \caption{Examples of slope error profiles at azimuthal angles of 0°, 90°, 180°, and 270° (left), and azimuthal dependence of the spread of the slope error distribution along the axial direction in the axial direction (right).}
  \label{fig:slperrhist_slperrspread}
\end{figure*}

\begin{figure*}[h!]
  \centering
  \includegraphics[width=18cm]{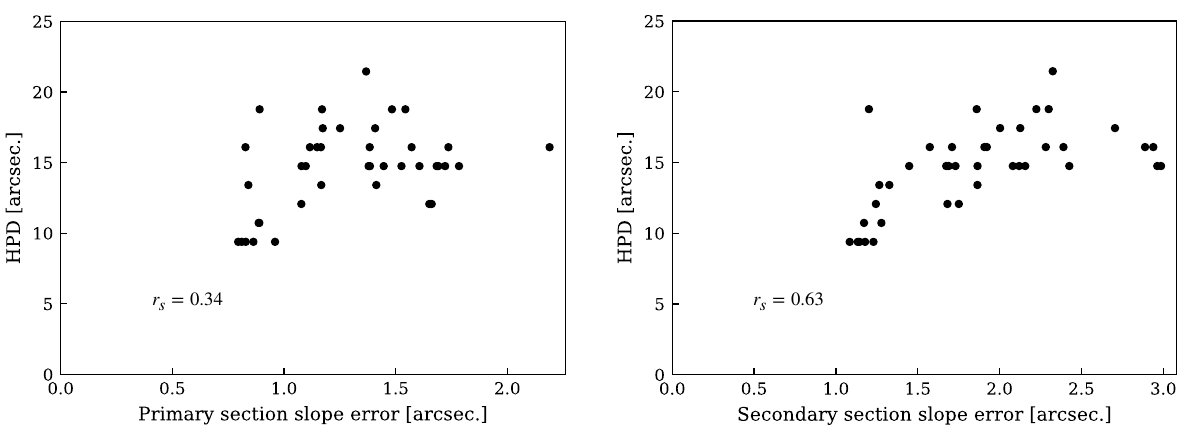}
  \caption{Correlation plots between angular resolution and the spread of the slope error distribution, shown for the primary section (left) and secondary section (right).}
  \label{fig:spotscan_corr}
\end{figure*}

The figure error data were measured with a sampling pitch of 0.05 mm along the axial direction and 1${}^\circ$ (approximately 0.5 mm) along the azimuthal direction\citep{Yamaguchi2023rsi}. 
Each spot-scan region corresponds, on average, to nine azimuthal data points, which were combined to produce a slope error histogram for that region. 
The slope error profile was obtained by differentiating the measured figure error data along the axial direction, and a 10 mm moving average window (5 mm at the edges) was applied to suppress high-frequency noise. 
Due to measurement limitations, the top and bottom 10 mm regions and the unpolished boundary zones near both ends (about 5 mm each) were excluded from the analysis. 
Examples of the slope error profiles are shown in Figure \ref{fig:slperrhist_slperrspread} left. 
The slope error amplitude tends to increase toward both edges, particularly in the secondary section, where deviations exceeding several tens of arcseconds were observed over scales longer than 10 mm. 
The peak structure near 0${}^\circ$ is likely caused by surface defects on the mandrel. Next, the slope error histograms were generated from these profiles, and their widths were used as an indicator of the angular deviation. 
Since the baseline component contributes exclusively to a shift in the focal length, it was subtracted prior to evaluation. 
For this purpose, the median value of the slope error distribution was defined as the baseline. 
The baseline slope errors of the primary and secondary sections were 0.13 arcsec and 0.87 arcsec, respectively, corresponding to a focal length shortening of about 1.3 mm, which is smaller than the depth of focus. 
The absolute median deviation from the baseline was then adopted as a measure of the slope error spread. 
The results are summarized in Table \ref{table:spotscan_hpd} and Figure \ref{fig:slperrhist_slperrspread} right. 
Both the primary and secondary sections show similar trends, though the secondary exhibits larger values overall. 
The smallest spreads (0.8 arcsec for the primary and 1.7 arcsec for the secondary) were found around the 200${}^\circ$--250${}^\circ$ azimuthal region, while the largest exceeded 3 arcsec near 0${}^\circ$. 
The Spearman’s rank correlation coefficient between the primary and secondary sections was 0.49, indicating a positive correlation that likely reflects the influence of common manufacturing factors such as replication stress or mandrel surface shape. 
We then compared the azimuthal dependence of the measured HPD values obtained from the spot-scan test with that of the slope error spread derived from the axial figure error analysis. 
Because the contribution of each section’s slope error to the two-reflection component depends on its individual surface profile, we simply examined the correlation for each section separately. 
As shown in Figure \ref{fig:spotscan_corr}, both sections exhibit a positive correlation between HPD and slope error width, with Spearman’s coefficients of 0.34 for the primary and 0.63 for the secondary. 
The stronger correlation in the secondary section is attributed to its generally larger slope error amplitude. 
These results suggest that axial figure errors of the mirror at least partially contribute to the observed degradation in angular resolution. 
Thus, improving the axial figure accuracy will be essential to further enhance the angular resolution in future electroformed mirrors.

\section{Summary and Conclusions} \label{sec:summary-conclusion}
We have developed a high-precision, 60-mm-diameter, full-circumference, double-reflection monolithic electroformed nickel mirror for space X-ray telescope mirrors, based on our electroforming technology originally established for millimeter-aperture, ultra-short-focal-length nanofocusing mirrors used in synchrotron X-ray microscopy. 
The detailed X-ray performance of this mirror was investigated both as a single mirror and after integration into the MMA.

To enable accurate characterization of such high-angular-resolution optics, we constructed HBX-KLAEES, a dedicated kilometer-long expanded-beam evaluation system at SPring-8 beamline BL29XUL. 
This system provides a high-brilliance, virtually point-like X-ray source with a very small diverging angle, which allowed us to measure both the sharp core and the large-angle components of the PSF with high fidelity.
Using HBX-KLAEES, the MMA achieved an angular resolution of 0.7 arcsec (FWHM) and 14 arcsec (HPD) at 12 keV, showing that the integration process preserved the intrinsic performance of the single mirror. 
These values represent the highest angular resolution achieved within the FOXSI series.
By comparing the X-ray results with the measured axial figure-error data, we identified a positive correlation between angular resolution and axial slope errors of both the primary and secondary sections. 
This demonstrates that the axial figure accuracy is a key factor limiting the achievable resolution of our mirrors.

Based on these performance results, the MMA incorporating this mirror was selected as one of the hard X-ray optics for the FOXSI-4 sounding-rocket mission, which was subsequently launched successfully.
More broadly, the electroformed-mirror fabrication framework forms a foundation for next-generation, high-angular-resolution X-ray optics for future space missions, as well as compact, ultra-short-focal-length systems suited for small-satellite platforms including CubeSats.

\begin{acknowledgments}
This work was supported by the Grants-in-Aid for Scientific Research (KAKENHI) from the Japan Society for the Promotion of Science (JSPS) under Grant Numbers JP22K18274, and JP20K20920 (IM), JP23H00156 (HM), and JP22H00134 and JP21KK0052 (NN).
Additional support was provided by JST SPRING (Grant Number JPMJSP2125) and the ISAS program for small-scale projects.
This work was also supported by the Iwadare Scholarship Foundation (RF and MY) and the Yokoyama Scholarship Foundation (RF).
KA acknowledges the generous support of the Hattori International Scholarship Foundation (HISF).
KS and KA also thank the “THERS Make New Standards Program for the Next Generation Researchers” for financial support.

\end{acknowledgments}
\bibliography{references_efoptics_fujii_pasp}{}
\bibliographystyle{aasjournalv7}



\end{document}